\documentclass[10pt]{article}
\usepackage[margin=0.6in]{geometry}
\usepackage[printwatermark]{xwatermark}
\usepackage{verbatim}
\usepackage{amsmath}
\usepackage{breqn}
\usepackage{amssymb}
\usepackage{multirow}
\usepackage{import}

\usepackage{xcolor}
\usepackage{float}
\usepackage{graphicx}

\usepackage[utf8]{inputenc}
\usepackage{gensymb}
\usepackage[section]{placeins}
\usepackage[numbers]{natbib}
\usepackage{lineno}
\usepackage{authblk}

\usepackage{titlesec}
\titleformat*{\section}{\normalsize\bfseries}
\titleformat*{\subsection}{\normalsize\bfseries}

\usepackage{setspace}
\usepackage{etoolbox}
\preto\longtable{\par\singlespacing}

\usepackage{hyperref}
\hypersetup{
	colorlinks = true,
	linkcolor = {red},
	citecolor = {blue},
	linkbordercolor = {white},
	citebordercolor = {purple}
}

\title{Constraining surface properties of asteroid (162173) Ryugu from numerical simulations of Hayabusa2 mission impact experiment} 

\author[1,*]{Martin Jutzi}
\author[1]{Sabina D. Raducan}
\author[2,3]{Yun Zhang}
\author[3]{Patrick Michel}
\author[4]{Masahiko Arakawa}
\affil[1]{Space Research and Planetary Sciences, Physikalisches Institut, University of Bern, Switzerland ($^*$E-mail: martin.jutzi@unibe.ch)}
\affil[2]{University of Maryland, College Park, USA.}
\affil[3]{Universite C\^ote d’Azur, Observatoire de la C\^ote d’Azur, Laboratoire Lagrange, Nice, France. }
\affil[4] {Graduate School of Science, Kobe University, Kobe 657-8501, Japan}

\begin{document}
\maketitle

%\linenumbers

% 
%%%%%%%%%%%%%%%%%%%%%%%%%%%%%%%%%%%
%The abstract — which should be no more than 150 words long and contain no references — should serve both as a general introduction to the topic and as a brief, non-technical summary of the main results and their implications.
% 

\centerline{\textbf{Abstract}}
The Hayabusa2 mission impact experiment on asteroid Ryugu created an unexpectedly large crater.  The associated regime of low-gravity, low-strength cratering remained largely unexplored so far, because these impact conditions cannot be re-created in laboratory experiments on Earth.
Here we show that the target cohesion may be very low and the impact probably occurred in the transitional cratering regime, between strength and gravity.  For such conditions, our numerical simulations are able to reproduce the outcome of the impact on Ryugu, including the effects of boulders originally located near the impact point. Consistent with most recent analysis of Ryugu and Bennu, cratering scaling-laws derived from our results suggest that surfaces of small asteroids must be very young. However, our results also show that the cratering efficiency can be strongly affected by the presence of a very small amount of cohesion. Consequently, the varying ages of different geological surface units on Ryugu may be due to the influence of cohesion.

%\linenumbers

\twocolumn

\section*{Introduction}\label{s:intro}

Both the projectile’s properties (e.g., mass, velocity), and the target’s properties (e.g., strength, gravity, porosity, structure) determine the size and morphology of impact craters on a planetary surface. Crater sizes and morphologies on the surface of asteroids can provide a direct diagnosis of their  material properties and their near-surface structure. Studies of the cratering process have important implications for understanding the chronology of the geological and geophysical evolution of asteroids, as well as for the design of planetary defence missions aimed at deflecting them by a kinetic impact.

The crater size is known to be controlled by gravity, strength, or a combination of the two, but it can also be controlled by porosity in the case of compaction-controlled craters \citep[][]{Housen2018}. Small asteroids like Ryugu have a very low gravity, but are also very weak, making it ambiguous which of these parameters controls the crater growth. While some mechanical properties of the surface of Ryugu can be inferred from the Hayabusa2 data, one large uncertainty is what is the surface strength. For the same impact energy, stronger surfaces are expected to have smaller craters than weaker surfaces. 

The small carry-on impactor (SCI) experiment on the surface of Ryugu had known impact conditions (i.e., projectile mass, impact velocity) and  represents an ideal test case to validate impact cratering models in the actual conditions of an impact onto an asteroid. Furthermore, it provides important insights to the properties, origin and evolution of the small body populations. The analysis of the cratering process induced by the recent SCI experiment on asteroid Ryugu concluded that the impact might have taken place in the gravity-dominated regime \citep{Arakawa2020} and therefore the surface cohesion on Ryugu must be very small ($<$ approximately 1 Pa). These conclusions were based on the extrapolation of laboratory-derived scaling relationships applied to the surprisingly large SCI crater size (with a diameter of about 14.5\,m \citep{Arakawa2020}), as well as on the behaviour of the ejecta curtain. 
However, because the Hayabusa2 impact experiment was the first of its kind, it is unclear whether traditional impact models and crater scaling-laws are applicable to the low-strength low-gravity cratering environment. How much the cratering process and its outcome, such as the crater size and morphology, were influenced by target surface inhomogeneities (i.e., the boulders located close to the impact point) is another important question. 

Using shock-physics codes, it has not been feasible so far to model the entire crater formation in the gravity-regime on small (100--1000\,m) asteroids, because of the vastly different time-scales of the shock-wave propagation and the crater formation.

Here we use the Bern Smoothed-Particle-Hydrodynamics (SPH) impact code \citep{Benz1995, Jutzi2008, Jutzi2015} to study the sensitivity of the SCI impact outcome to target cohesion and target inhomogeneities (i.e., boulders). The late-stage evolution of the crater is simulated using a recently developed fast integration scheme. Using this approach, the outcome of the SCI impact on Ryugu can be well reproduced. Our simulations show that the cratering efficiency in the investigated low-strength low-gravity cratering environment is very large, but can be strongly affected by the presence of a small amount of cohesion.

\section*{Results}
The shock physics code Bern SPH includes material models suitable to simulate the behaviour of geological materials, various equations of state and a porosity compaction model.
%To simulate the late-stage evolution, we apply a recently developed numerical approach \citep[][]{Raducan2022} that allows for faster calculation times and the simulation of the entire process (see Methods, subsection Modelling approach for the late stage evolution). This was achieved by applying a transition to a low-speed medium in the shock physics code at some advanced time of the simulation.
We simulate the late-stage evolution by applying the numerical approach developed in \citep[][]{Raducan2022} as it allows for faster calculation times and the simulation of the entire process (see Methods, subsection Modelling approach for the late stage evolution). This is achieved via application of a transition to a low-speed medium at some advanced time of the simulation.
In fact, after the initial shock has passed, the late-stage evolution is governed by low-velocity granular flows, which can be accurately modelled using a low bulk sound speed material (i.e., a material with a small $c_s$, allowing for a larger time step $dt$). This approach has been extensively tested and validated against laboratory experiments into homogeneous and heterogeneous targets \citep[][]{Ormo2022}. Furthermore, the comparison with an alternative, SPH-SSDEM modelling scheme, in which the late stage crater evolution is modelled using the soft-sphere discrete-element method (SSDEM) to directly simulate the granular material dynamics, shows a very good agreement (see Methods, subsection Modelling approach for the late stage evolution). 

%%%%%%%%%%%%%%%%%%%%%%%%%%%%%%%%%%%%%%%%%%%%%%%%%%%%%%%%%%%%
The target material response to shear deformation is described by a simple pressure-dependent strength model \citep{Lundborg1967, Collins2004}. A measure of strength is given by the material's ability to withstand different types of stress states. A granular material, for example, can still have a significant amount of shear strength originating from van der Waals forces and the inability of the interlocking particles to move apart and slide over one another \citep{Scheeres2010}. Here we study the effects of the shear strength at zero pressure, which is often referred to as cohesion. 
Since the surface strength is the unknown quantity in the study, we first varied the initial cohesion ($Y_0$ = 0--0.5 Pa) and coefficient of friction ($f$ = 0.4--1.0) (see Methods, subsection Shock physics code model). For our nominal simulations, we keep the target initial density constant at 1300 kg/m$^3$ (initial porosity $\phi_0$ = 50\%), which is slightly higher than the bulk density of Ryugu (1190 kg/m$^3$) measured by Hayabusa2 \citep{Sugita:2019sh}. We include 50\%, microporosity (as described in \cite{Jutzi2008, Jutzi2010}), which is consistent with recent analysis of Hayabusa2 samples returned from Ryugu \citep{Yada2021}.
 
%%%%%%%%%%%%%%%%%%

\begin{figure}[h]
\centering
\includegraphics[width=\linewidth]{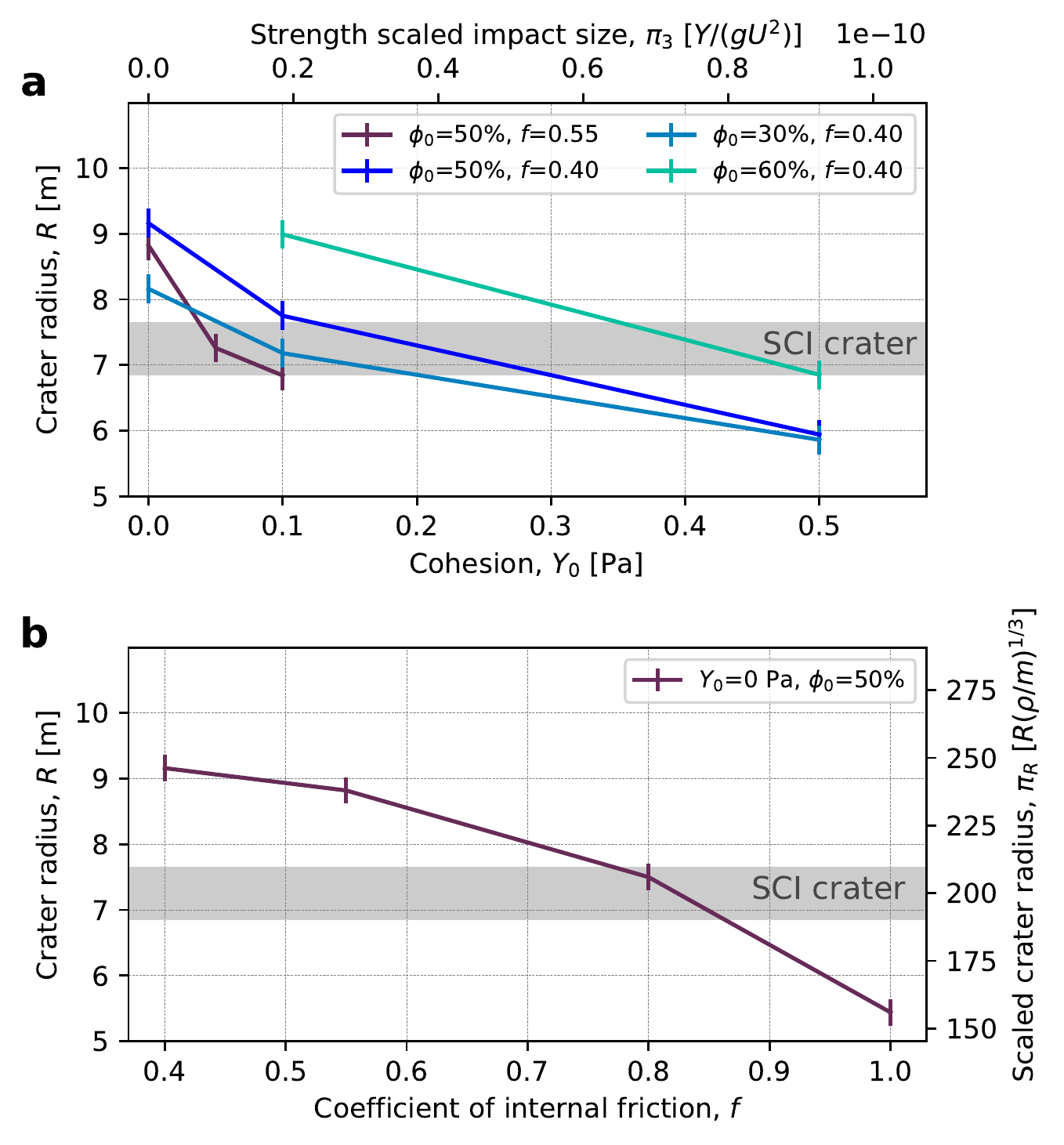}
\caption{Crater radius as a function of material properties. a) Crater radius as a function of cohesion $Y_0$ for different friction coefficients ($f$ = 0.4 and $f$ = 0.55) and porosities (30, 50, 60 \%). b) Crater radius as a function of the coefficient of internal friction for 50\% porous, cohesionless targets ($Y_0$ = 0). The horizontal line shows the measured SCI crater size at the pre-impact level (7.25$\pm$0.8\,m), with its associated error \citep{Arakawa2020}. Source data are provided as a Source Data file.}
\label{fig:rcrat_f_c}
\end{figure}

Our simulations show that the resulting final crater is very sensitive to our choice of cohesion and coefficient of internal friction (Figure \ref{fig:rcrat_f_c}). The solution is not unique and the same crater size can be produced by impacting targets with a combination of mechanical properties (represented by the friction coefficient and the cohesion). In order to match the observed size of the SCI crater, a cohesionless ($Y_0$ = 0 Pa) target with the nominal porosity of 50\% requires a relatively high friction coefficient ($f>$ 0.8).
On the other hand, the required crater size is also produced with a cohesion of about 0.2 Pa and a small friction coefficient ($f$ = 0.4). 

%%%%%%%%%%%%%%%%%%%%%%%%%%%%%%%%%%%%%%%%%%%%%%%%%%%%%%%%%%%
It is not possible to constrain the cohesion at the required level, i.e., to distinguish between 0.0 Pa and 0.2 Pa cohesion, based on the available observations. However, reasonable assumptions can be made regarding the friction coefficient. 
The angle of repose of cohesionless materials has been measured to be 22$^\circ$ for glass beads \citep[e.g.][]{Lajeunesse:2005,Chourey:2020} and 30$^\circ$ for quartz sand \citep[e.g.][]{Lube:2004,Chourey:2020}. For lunar regolith simulant with a small but significant cohesion in the range of 0.5--2 kPa, the angle of repose is significantly higher, in the range of  35--45$^\circ$ \citep[e.g.][]{Mitchell:1972,Chourey:2020}. The corresponding friction coefficients are about $f$ = 0.4 (glass beads), $f$ = 0.55 (quartz sand) and $f$ = 0.7--0.9 (regolith simulant) (Methods, subsection Shock physics code model). 
We consider cohesionless quartz sand as the best representative of Ryugu's subsurface material because of the evidence for a very low cohesion ($<$ 1 Pa), and we define $f=0.55$ as nominal friction coefficient. The friction coefficient for glass beads is regarded as a lower limit for geological materials and our results for targets with $f$ = 0.4 and 50$\%$ porosity suggest that the cohesion present on Ryugu at the SCI impact site cannot be higher than approximately 0.2 Pa (Figure \ref{fig:rcrat_f_c}a).

\begin{figure*}[h]
\centering
\includegraphics[width=0.7\linewidth]{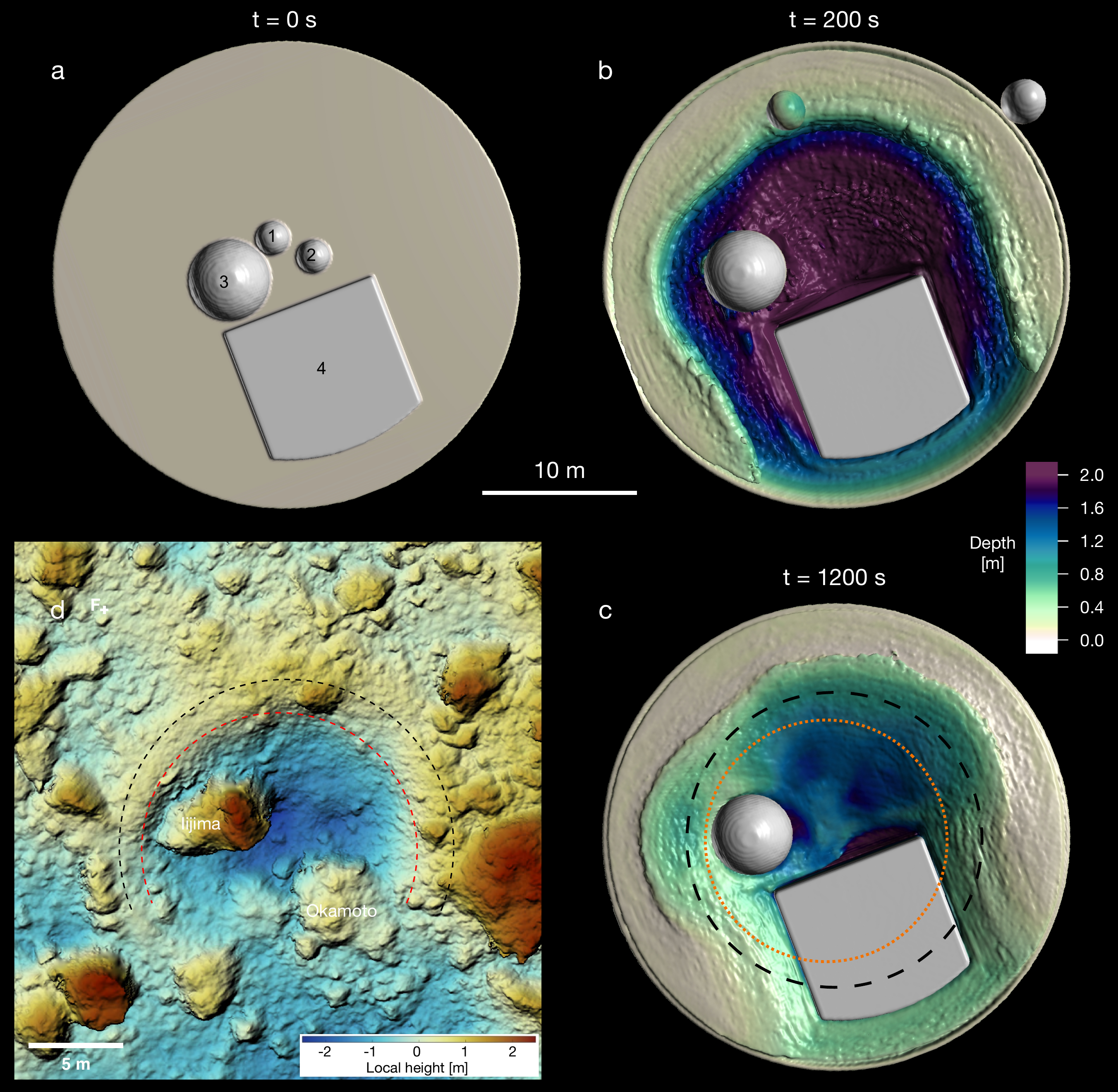}
\caption{End-to-end SPH simulation of the SCI impact. a) -- c) Snapshots of the simulation are shown at different times. At $t$ = 1200s the crater evolution is finished. d) Observed SCI crater on Ryugu \citep[][]{Arakawa2020}. The local height is measured relative to point F. The crater dimensions are also indicated by the dotted orange (crater wall) and dashed black (crater rim) lines in c). The overall outcome in terms of crater size and boulder displacement is well reproduced. In the simulation, the largest block (4) is moved by 10--20 cm and the medium boulder (3) by 2 m, which is roughly consistent with the observed displacement of the large boulders at the SCI impact site (36 cm for the large Okamoto boulder; 2.58 m for the medium sized Iijima boulder; \cite{Honda:2021}). In the simulation, the small boulders (1,2) are ejected out of the crater, as observed in the case of the SCI impact for some of the boulders (e.g. boulders \textit{mb01} and \textit{mb02}; \cite{Honda:2021}).}
\label{fig:surface_snapshots}
\end{figure*}

\begin{figure*}[h]
\centering
\includegraphics[width=0.75\linewidth]{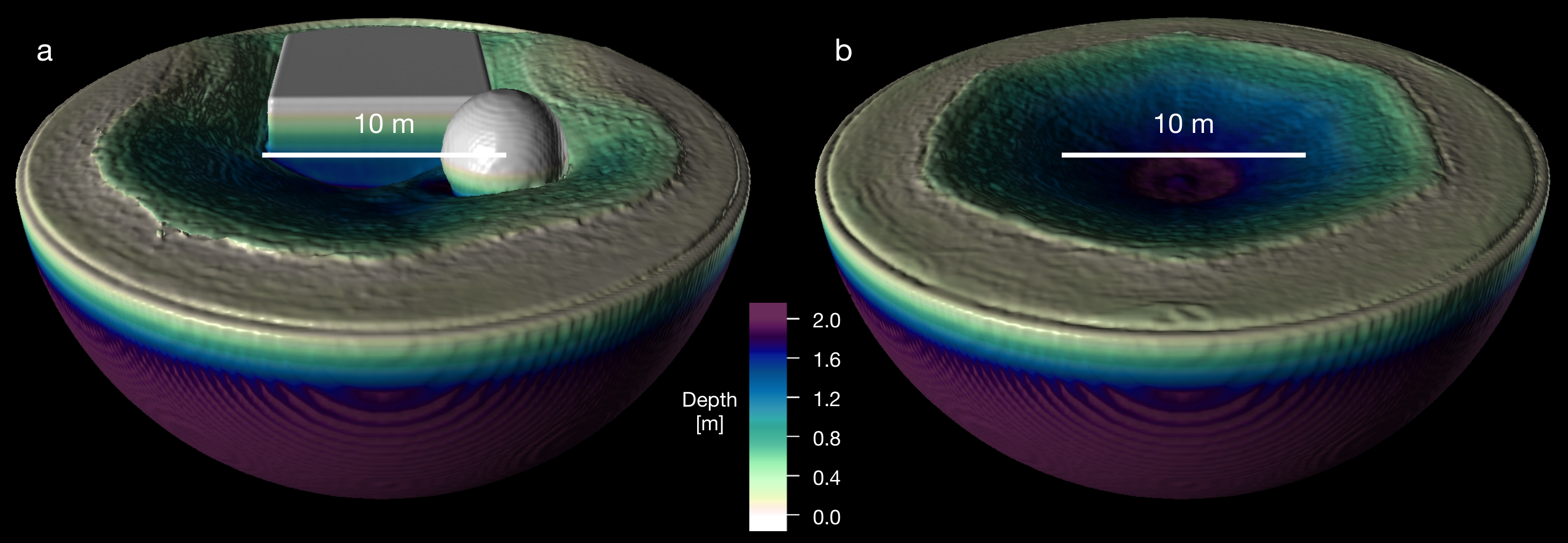}
\caption{Comparison of SCI impact simulations with varying target surface properties. a) Targets with boulders and b) targets without boulders, with otherwise the same impact conditions. In both cases, the homogeneous target material is cohesionless (while the boulders have significant cohesion). The case shown in a) corresponds to the one shown in Figure 1. The crater radius in the case of the homogeneous target is 8.82 m. }
\label{fig:surface_comparison}
\end{figure*}

However, in addition to friction and cohesion, the target porosity is also an important parameter \citep[e.g.,][]{Housen2018}. The relation between porosity and cohesion depends on the scale and type of porosity. For instance, a target made of (porous) boulders or small particles can have a relatively high porosity, but an overall small bulk cohesion (contact forces between the particles). Such targets would have a bulk crushing strength much larger than the bulk cohesion.

For a surface with a very low bulk cohesion ($\lessapprox$ 1 Pa) and significant crushing strength ($>>$ 1 Pa), the SCI impact is not expected to be in the so-called compaction cratering regime (in which the crater forms mostly by crushing and permanent compaction of the target \citep[][]{Housen2018}). Our results also confirm this, even for highly porous targets (i.e., the amount of compacted material in the final crater is negligible). This is because of the relatively small scale of the SCI impact. In craters formed at small scales, outward expansion caused by the late-time material flow dominates over compaction \citep[][]{Housen2018}. The crater would form in the compaction cratering regime only if the compressive strength was of the same order as the cohesion ($\lessapprox$ 1 Pa), while weak soils have compressive strength $>$ 10$^4$ Pa \citep[][]{Housen2003}.
However, porosity is still relevant for the crater size, as it affects the coupling to the target and it also determines the bulk density, which is an important parameter for the cratering efficiency in both the strength and gravity dominated cratering regimes \cite[e.g.][]{Holsapple1993}.
In order to investigate the effect of porosity and to obtain an upper limit of cohesion on Ryugu's surface, we perform impact simulations into targets with $f$=0.4, varying cohesion ($Y_0$ = 0-0.5 Pa) and varying initial porosity ($\phi_0$ = 30-60\%). Including these porosity variations, our results suggest an overall upper limit for the cohesion on Ryugu's surface of approximately 0.5 Pa (Figure \ref{fig:rcrat_f_c}a).

For a subset of our simulations, boulders with varying sizes are distributed within the target (Fig.~\ref{fig:surface_snapshots}), in a similar way as that observed at the SCI impact zone. The boulders have the same density as the matrix material but a cohesion of 100 MPa. The boulders are assigned a high strength because we are interested in the movement of intact boulders rather than their fragmentation (fragmentation has not been observed in the SCI impact). The homogeneous matrix material is considered to be cohesionless ($Y_0$ = 0\,Pa) and with a coefficient of friction $f$ = 0.55. The sizes and positions of these boulders (see Methods, subsection Boulder position and impact geometry) are representative of the large, medium and small boulders present close to the SCI impact point \cite{Honda:2021} on Ryugu.

The overall outcome of the SCI impact, including the displacement of the boulders (Fig.~\ref{fig:surface_snapshots}), is well reproduced in our simulations. The presence of the large block leads to an asymmetric crater similar to the observed one. The comparison to a simulation without boulders but otherwise the same target properties (Fig.~\ref{fig:surface_comparison}) shows that the presence of the boulders within the target affects the resulting crater size only by a minimal amount ($<$ 5\%). In the simulations, the transient crater is reached at about $t \approx$ 200-300 s (Fig.~\ref{fig:surface_snapshots}), which is consistent with the observations \citep{Arakawa2020}.

Estimates of an asteroid's surface age rely on the relationship between the projectile's size and the crater's size, which can be derived from the crater scaling-laws. The general form of the scaling relationship for the crater radius, $R$, depends on velocity, $U$; radius, $a$; density, $\delta$; mass, $m$ of the impactor and the target properties: density, $\rho$; strength, $Y$, and gravitational acceleration, $g$. It is given by \cite[e.g.][]{Holsapple1993,Raducan2020}: 

\begin{equation}\label{eq:pi_R}
\begin{split}
R\left(\frac{\rho}{m}\right)^{1/3} = K_{R1}\Bigg\{\frac{ga}{U^2} \left(\frac{\rho}{\delta}\right)^{\frac{(6\nu-2-\mu)}{3\mu}}  \\
+\left[K_{R2}\left(\frac{Y}{\rho U^2}\right)\left(\frac{\rho}{\delta}\right)^{\frac{(6\nu-2)}{3\mu}}\right]^{\frac{2+\mu}{2}}\Bigg\}^{\frac{-\mu}{2+\mu}}
\end{split}
\end{equation}
where the constants $\mu$, $\nu$, $K_{R1}$ and $K_{R2}$ are determined empirically \cite[e.g.][]{Raducan2020}. 

%+\left[K_{R2}\left(\frac{Y}{\rho U^2}\right)\left(\frac{\rho}{\delta}\right)^{\frac{(6\nu-2)}{3\mu}}\right]^{\frac{2+\mu}{2}}\right\}\Bigg\}^{\frac{-\mu}{2+\mu}}

Using the nominal coefficient of friction ($f$ = 0.55), we investigate the dependence of the impact outcome to the impactor size for different values of cohesion. The simulation results are then used to  calibrate Eq.~\eqref{eq:pi_R}, which yields the fitting constants $K_{1}$ = 0.60, $K_{2}$  = 0.25, $\mu$ = 0.44 and $Y=Y_0$ = 0.05 Pa. 

Our analysis shows that even in this low-strength low-gravity regime - which is not accessible by laboratory experiments for the size and timescales relevant to asteroid cratering - the general form of the impact cratering scaling (Eq.~\ref{eq:pi_R}) allows for a very good fit of the crater sizes as a function of the impactor size and target strength (i.e., cohesion).
Our results also suggest that a cohesion of 0.05 Pa is required to match the SCI crater size for nominal material parameters. This means that the SCI impact may have taken place at the transition between the strength and gravity regime (Fig.~\ref{fig:scaling}), where cohesive and gravitational forces determine the impact outcome concurrently. 

\section*{Discussion}

The cratering efficiency defined as $\Pi_R=R\left(\frac{\rho}{m}\right)^{1/3}=D_{crat}/D_{proj}\left(\frac{3\rho}{4 \pi \delta}\right)^{1/3}$ (left term in Eq.~\eqref{eq:pi_R}), where $D_{crat}$ is the crater diameter and $D_{proj}$ is the projectile diameter, largely determines the surface ages of asteroids based on crater counting. 

\begin{figure}[h!]
\centering
\includegraphics[width=\linewidth]{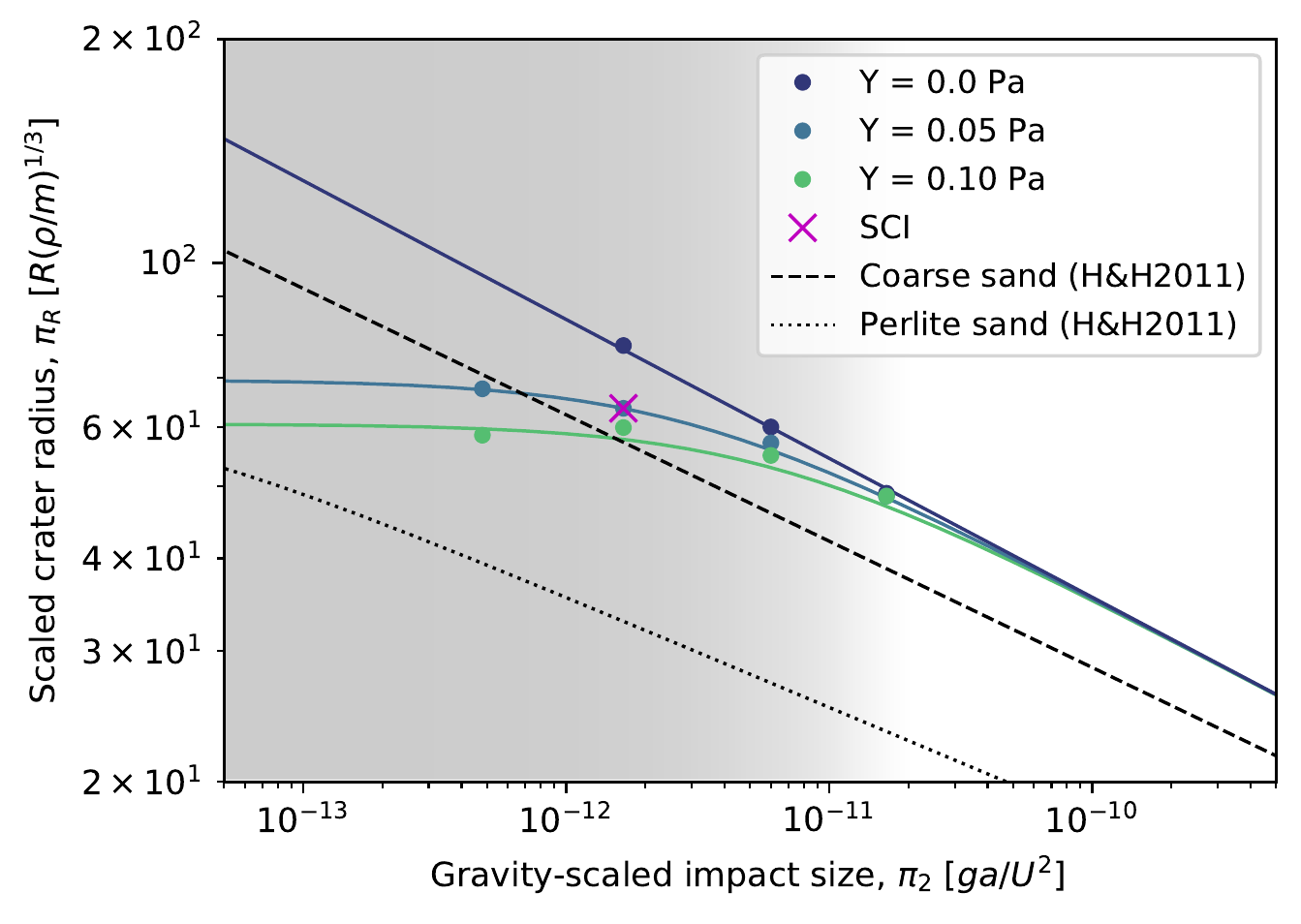}
\caption{Impact outcome as a function of impact size. Shown is the scaled crater radius as a function of gravity scaled impact size for different target cohesions. Shown are the results of the simulations using a nominal coefficient of friction of $f = 0.55$. The scaling relationships for coarse and perlite sand \citep[H$\&$H2011;][]{Housen2011} are shown for comparison. The case with a cohesion of $Y$ = 0.05 Pa leads to the best fit with the observed SCI crater. Even small increases of the cohesion lead to significantly smaller craters. The background gradient shows the transition between the strength and gravity dominated cratering regimes. Source data are provided as a Source Data file.}
\label{fig:scaling}
\end{figure}

As shown by the SCI experiment and confirmed by the results of the study presented here, small-scale impacts in low-cohesion targets have a very large cratering efficiency, with a correspondingly large ratio of $D_{crat}/D_{proj}$ of $>$ 100 (Fig.~\ref{fig:scaling}; with $D_{crat}/D_{proj}=\Pi_R \times\left(\frac{3 \rho}{4 \pi \delta}\right)^{-1/3} \approx \Pi_R \times 2.1 $). This is in contrast with the assumption of a (constant) small ratio of $D_{crat}/D_{proj}=10$ that has been suggested to be valid over a larger range of scales \citep{Bottke2020}. 
Important consequences of such a high cratering efficiency are high resurfacing rates and young crater retention ages.
For example, a ratio $D_{crat}/D_{proj}$ of about 100 leads to an about $10^3$ times higher crater formation rate (assuming $q=-4$ for the slope of the impactor population, see Methods, subsection Calculations of the impact frequencies)  than a ratio of $D_{crat}/D_{proj}=10$ and a correspondingly younger surface age. 
There is increasing evidence that the surfaces of both recently visited asteroids Ryugu and Bennu are very weak and our results suggest that their surface must therefore be very young (of the order of approximately 1--10 Myr). These findings are consistent with most recent analysis \citep[e.g.][]{cho:2021,Bierhaus2022,Perry2022}, but in contrast to previous studies which proposed much older surface ages (approximately 1 Gyr) based on low cratering efficiencies, corresponding to strength regime scaling \citep[e.g.][]{Walsh2019,Bottke2020}.

%Absolute model ages of each geologic unit on Ryugu have been shown to be in the range of about 2-30 Myrs \citep{cho:2021}, assuming that the cohesion of Ryugu is negligible and the gravity controls the crater formation. 
Assuming asteroid Ryugu's cohesion is negligible and crater formation is controlled by the gravity, \cite{cho:2021} showed that the absolute model ages of asteroid Ryugu's geologic units are in about 2-30 Myrs range.
As also indicated by the observations, there are variations in surface properties (e.g., strength), which possibly lead to varying ages for the different units. Such differences in surface ages might indeed be explained by small variations in cohesion, which lead to considerable changes of the cratering efficiencies resulting from the crater scaling-laws used here. 
As an example, our best fit scaling-law (Fig.~\ref{fig:scaling}; $Y$ = 0.05 Pa) leads to $D_{crat}/D_{proj}$ of about 180 for an impactor size of $a=0.1$ m, which in turns leads to a crater radius of about 18 m (assuming an impact velocity of 5 km/s). Using a slightly higher cohesion of 1 Pa, we obtain $D_{crat}/D_{proj}$ of about 110 and a crater radius of about 11 m. Following the procedure used in \citep{Jutzi2019}, this leads to an approximately 4 times higher rate of crater formation in this size range for the $Y$ = 0.05 Pa surface compared to the $Y$ = 1 Pa surface (see Methods, subsection Calculations of the impact frequencies) and a correspondingly younger surface age. As illustrated by this example, even a small amount of cohesion leads to a significantly reduced cratering efficiency. The influence of cohesion becomes larger with decreasing size scale (Fig.~\ref{fig:scaling}) and leads to a crater reduction effect for small-scale impacts on low-strength asteroid surfaces. Recent studies have shown that on weak asteroids, resurfacing by seismic shaking is not efficient \citep{Takaki:2022}.
On the other hand, armouring by surface boulders has been proposed to explain the paucity of craters smaller than 2-3 m on the asteroid Bennu \citep{Bierhaus2022}. Here, we demonstrate that the presence of a small amount of cohesion can actually contribute significantly to the crater reduction effect and therefore to the depletion of small craters in the observed crater distribution on small asteroids \citep[e.g.,][]{Tatsumi:2018, Takaki:2022}.

%%%%%%%%%
\newpage 
\onecolumn
\appendix

\section*{\large{Methods}}
\subsection*{Shock physics code model}
We modelled the target using the Tillotson equation of state (EoS) for basalt \citep{Benz1999}, with a modified initial bulk modulus (see next section) for the simulations of the late stage evolution. 
For the target material response to shear deformation we applied a simple pressure-dependent strength model, typical of geological materials \citep{Lundborg1967, Collins2004}. The Lundborg strength model describes the yield strength as:
\begin{equation}\label{eq:yield}
    Y = Y_0 + \frac{fP}{1+fP/(Y_{dm}-Y_0)},
\end{equation}
where $P$ is pressure, $f$ is the coefficient of internal friction, $Y_{dm}$ is the limiting strength at high pressure and $Y_0$ is the cohesion. 

We use a constant cohesion, $Y_0$, for the weak asteroid materials considered in this study and apply a strain-based weakening model that prevents artificial clumping (similar to the approach used in \cite{Collins2008}). Our model uses a linear relation between cohesion $Y_0$ and total strain $\epsilon_{tot}$, and it is assumed that for $\epsilon_{tot} \geq 1$ cohesion is lost. 
Instead of using a tensile fragmentation model, like it was implemented in previous studies \citep[e.g.,][]{Jutzi2015}, the tensile strength is defined by extrapolating the yield strength (versus pressure) curve (Eq. \ref{eq:yield}) to intersect the pressure axis. In addition, the pressure is limited to a minimum value $P_{min} \geq -Y_0$.

To compute the coefficient of internal friction from the measured angle of response $\theta$ \citep{Chourey:2020,Mitchell:1972}, we follow the relation described in \cite{Luther2022}: 
\begin{equation}
    f=\frac{2 \sqrt{2} sin(\theta)}{3-sin(\theta)}
\end{equation}

The porosity was modelled using the $P-\alpha$ model \citep{Jutzi2008} with a crush curve defined by the parameters $P_s$, $P_e$, $P_t$, $\alpha_t$, $n_1$, $n_2$ and initial distension, $\alpha_0$.
$P_e$ is the pressure defining the transition between the elastic and plastic regimes, $P_s$ the pressure at which all pores are compacted (distention $\alpha$ = 1) and the parameters $P_e$ $<$ $P_t$ $<$ $P_s$ and 1 $<$ $\alpha_t$ $<$ $\alpha_0$ indicate a transition between two regimes with different slopes ($n_1$ and $n_2$).

For both target and projectile materials, the Tillotson EOS is applied. For the relatively low SCI impact velocity, no significant vaporisation or melting is expected and the Tillotson EOS is appropriate for this impact regime \citep{Raducan2022b}.  
The relevant material parameters used in this study are summarised in Table~\ref{table:model_parameters}.

\begin{table}[h]
 \centering 
    \footnotesize
	\caption{Material model parameters for impact simulations into Ryugu analogues.}
    \vspace{0.3cm}
	\begin{tabular}{l@{\hskip 0.1in}l@{\hskip 0.1in}l@{\hskip 0.1in}l}
    Description        & Impactor & Homogeneous material & Boulders  \\
    \hline
    Material             & Aluminium & Basalt & Basalt \\
    Impact angle ($^o$)  & 90 & -- & --   \\
    Impact speed (km/s)  & 2 & --     & --     \\
	\hline
	Equation of state        & Tillotson$^{a}$ & Tillotson$^b$ & Tillotson$^b$ \\
	Strength model           & von Mises & LUND$^c$ & LUND$^c$ \\
	\hline
	LUND strength parameters$^c$ \\
	Damage strength at zero pressure, Y$_0$ (Pa)   & --   & 0/0.05/0.1/0.5  & $1\times10^8$ \\
	Strength at infinite pressure, Y$_{\inf}$ (GPa) & 0.34  & 3.5  & 3.5    \\
	Internal friction coefficient (damaged), $f$    & --  & 0.4/0.55/0.8/1.0 & 0.8\\
	\hline
	Porosity model parameters$^d$           \\ 
	Initial porosity, $\phi_0$             & -- & 50\% & -- \\
	Initial distension, $\alpha_0$         & -- & 2 & -- \\
	Transition distention, $\alpha_t$                             & -- & 1.25 & -- \\
	Pressure at full compaction, $P_s$ (MPa)                            & -- & 213 & -- \\
	Transition pressure, $P_t$ (MPa)                            & -- & 68 & -- \\
	Elastic pressure, $P_e$ (MPa)                            & -- & 1.0 & --  \\
	Slope regime 1, $n_1$                                  & -- & 12 & --  \\
	Slope regime 2, $n_2$                                  & -- & 3 & --  \\
	\hline
    \multicolumn{3}{l}{
    $^a$\cite{Tillotson1962};
    $^b$\cite{Benz1999};
    $^c$\cite{Lundborg1967};
    $^d$\cite{Jutzi2008}.}
	\end{tabular}
	\label{table:model_parameters}
\end{table}
\FloatBarrier

\newpage
\subsection*{Initial conditions}
The target is modelled as a hemispherical domain with a radius of 15 m (Fig.~\ref{fig:surface_comparison}) using approximately $10^7$ SPH particles. We use a spherical  projectile (Table~\ref{table:model_parameters}) with a mass of 2 kg, impacting head-on. Recent studies \citep{Raducan2022b} have shown that for low-strength targets with (with associated high cratering efficiencies), the projectile shape does not significantly influence the crater size. Because of the  large simulation domain required to model the impacts with very high cratering efficiencies, the projectile shape cannot be resolved in our simulations. We  approximate the SCI impactor using a homogeneous sphere with a the same diameter as the SCI at impact but with a correspondingly lower density (roughly corresponding to Aluminium), following the approach used in previous studies \citep[e.g.][]{Raducan2019,Raducan2022b,Jutzi2014}. 

For the simulations with varying projectiles (Fig.~\ref{fig:scaling}), we use projectile masses of 0.05, 2, 100 and 2000 kg, respectively, and adapt the size of the computational domain and the computation times accordingly. 

All models used a constant gravity acceleration of $1.4 \times 10^{-4}$ m/s$^2$. 
\subsection*{Modelling approach for the late stage evolution}

Modelling the entire crater formation in the gravity-regime on small (about 100\,m -- 1000\,m) asteroids is challenging because of the vastly different time-scales of the shock-wave propagation and the crater formation. The first time-scale is governed by the elastic wave velocity $c_s$, typically a few km/s for rocks. For example, to model a SCI-like impact and to ensure numerical stability, the so-called Courant criteria requires that the timestep, $dt$, is smaller than the simulation resolution divided by the sound speed in the target, $c_s$: $dt < \mathrm{resolution} /c_s \simeq 10^{-6} - 10^{-7}$\,s. On the other hand, the crater formation time in the gravity regime can last up to few hundred seconds \citep[][]{Arakawa2020}. 

\subsubsection*{Direct shock physics code modeling of late stage evolution}
To model the late-stage evolution of the crater grow we follow the approach introduced in \cite{Raducan2022} and apply in the shock physics code calculation a transition to a low-sound-speed medium. After the initial shock has passed, the late-stage evolution is governed by low-velocity granular flow, which can be accurately modelled using a low bulk sound speed material (i.e., a material with a small $c_s$, allowing for a larger $dt$).

At this time  $t_{transition}$ we switch to a low-sound-speed medium (i.e., a fast scheme), which allows us to model the late-stage evolution up to about a thousand seconds after the impact. In this step, we apply a simplified Tillotson EoS, in which all energy related terms are set to zero. The remaining leading term of the EoS is governed by the bulk modulus $P= A (1-\rho/\rho_0)$, which also determines the magnitude of the sound speed. We use $A\approx0.027$\,MPa and also reduce the shear modulus proportionally. 

This approach has been extensively tested and validated against laboratory experiments into homogeneous and heterogeneous targets \citep[][]{Ormo2022}.

The transition time $t_{transition}$ depends on the scale of the impact. We use the following definition $t_{transition}\simeq L/v_{sound}\times 10$ where $L$ is the length scale (chosen to be the size of the computational domain) and $v_{sound}$ the sound speed of the material. According to this definition,  $t_{transition}$ corresponds to roughly 10 sound crossing times. For the SCI-scale impact, we use $t_{transition}$ = 0.1 s. 

To test the robustness of the approach, we performed additional simulations with different transition times. For this test, we use the larger scale impactor (10 $\times$ SCI-size; Fig.~\ref{fig:scaling}) as the simulations are faster to perform at larger scales. We use transition times of $t_{transition}$ = 0.5, 0.8, 1.4 and 2.0 s.  For these different transition times, the resulting crater radii of 52.0, 52.8, 52.8 and 53.6 m agree very well with each other; the deviations are are within the uncertainty of the crater size determination ($\pm$ 1.1 m).

\subsubsection*{SPH-SSDEM framework for late stage evolution}
An alternative approach to direct SPH simulations is to use a SPH-SSDEM framework \citep{Zhang2021}, in which the late stage crater evolution is modelled using the soft-sphere discrete-element method (SSDEM).
The SSDEM method can be used to model low-velocity collisions and the flow of granular materials and it is well suited to simulate the late-stage of the impact process. For this study, the parallelized hierarchical tree SSDEM code, pkdgrav, is used.
%A granular physics model including 4 dissipation/friction components in the normal, tangential, rolling, and twisting directions is applied for computing particle contact interactions and controlling the material shear strength \citep{Schwartz:2012,Zhang2017}.  
To compute particle contact interaction and control the material shear strength, we used a granular physics model with four dissipation/friction components in the normal, tangential, rolling and twisting directions \citep{Schwartz:2012,Zhang2017}.

The combination of the SPH and SSDEM methods has been used in previous studies to simulate small body impact processes \citep[e.g.][]{Schwartz2018,Ballouz:2019}. The initial shock propagation and fragmentation stage is simulated using SPH and the outcome is then transferred into the SSDEM code, which computes that late-stage evolution. The particle-based description of the two methods allows direct information transition at the particle level. However, there are several issues with this direct particle-particle transition, for instance the handling of particle overlaps and maintaining linear and angular momentum constant \citep[][]{Ballouz:2019}.
%Furthermore, the direct transition with comparable particle resolution in the subsequent SSDEM simulations leads to unachievable computational time to complete the crater growth modelling \citep[][]{Schwartz2016}, since SPH simulations use very high particle resolution for properly solving the shock wave propagation and energy dissipation.  
Additionally, because SPH simulations use very high particle resolution for properly solving the shock wave propagation and energy dissipation, the direct transition with comparable particle resolution in the subsequent SSDEM simulations to complete the crater growth modelling would require computational times that are not achievable \citep[][]{Schwartz2016}.

We use a shape-construction algorithm \citep[][]{Ballouz:2019} for the velocity field SPH-SSDEM transition procedure to solve these issues. 
%The procedure consists of four steps: (1) using the SSDEM with a boundary geometry, gravity conditions, and material parameters that are consistent with the SPH simulation, a granular bed with a predefined particle size distribution is produced and settled down; (2) based on the given SPH output, we use the $\alpha$-shape-construction algorithm to construct a surface that isolates the compact material from the fast-moving ejecta whose speeds exceed a given limit; (3) the isolated surface is then used to carve out the surface of the SSDEM granular bed, and a nearest neighbour search is conducted to map the linear and angular momentum from the SPH compact material data to the SSDEM bed; (4) finally, the fast-moving ejecta is added into the SSDEM simulation scenario as individual particles. 
The procedure consists of four steps: (1) we prepare a granular bed with a predefined particle size distribution in which material parameters are consistent with the SPH simulation, and let the particles settle down in the SSDEM boundary geometry and gravity conditions, (2) based on the given SPH output, we use the $\alpha$-shape-construction algorithm to construct a surface isolating the compact material from the fast-moving ejecta whose speeds exceed a given limit; (3) we then use the isolated surface to carve out the SSDEM granular bed surface, and conduct a nearest neighbour search to map the linear and angular momentum from the SPH compact material data to the SSDEM bed; (4) finally, we add fast-moving ejecta into the SSDEM simulation scenario as individual particles.

\begin{figure}[h!]
\begin{center}
\includegraphics[width=8cm]{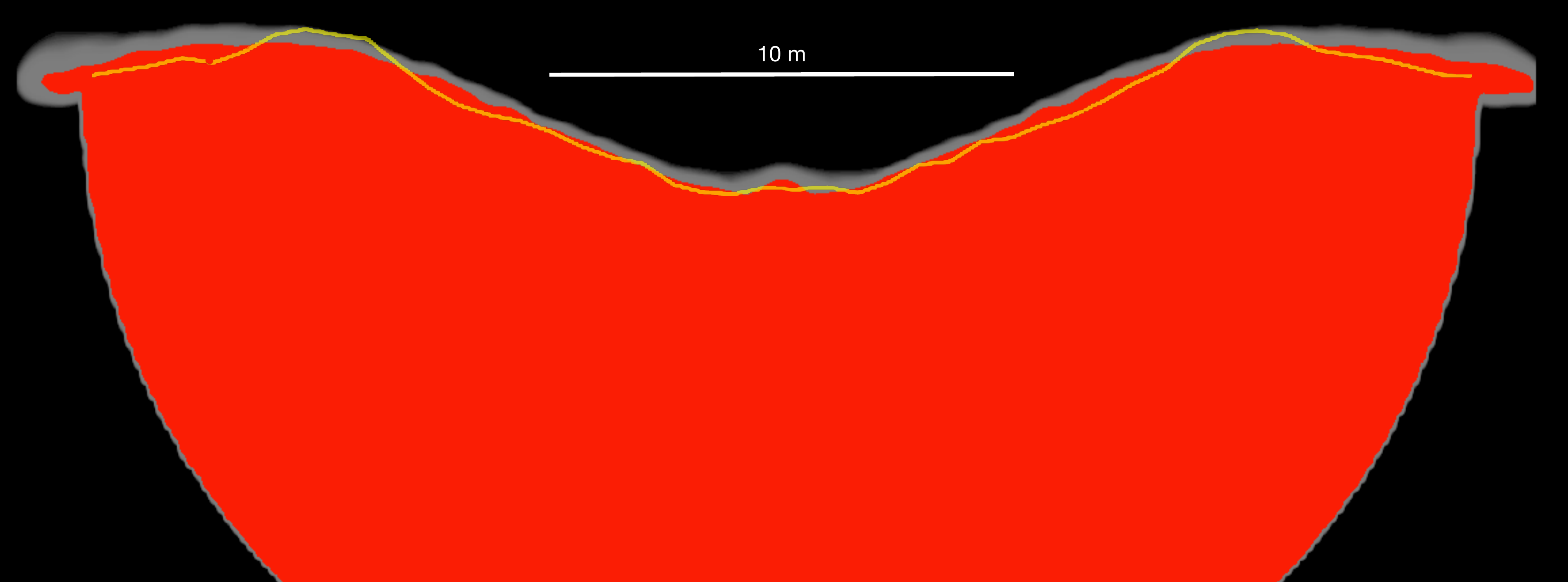}
\caption{Comparison of two numerical approaches. Shown are the results of direct SPH (this work) and SPH-SSDEM simulations \citep{Zhang2021}. The crater profile obtained by the hybrid SPH-SSDEM simulation (yellow line) is over-plotted on the results of the direct SPH simulation. Indicated in gray is the uncertainty due to the SPH kernel interpolation and associated smoothing length. The final crater size and morphology are in good agreement.} 
\label{fig:sph-ssdem}
\end{center}
\end{figure}

Initial simulations using the SPH-SSDEM framework show a very good agreement with the outcome of the direct SPH shock physics code calculation (Fig.~\ref{fig:sph-ssdem}).

\subsection*{Boulder position and impact geometry}
\begin{figure}[h!]
\begin{center}
\includegraphics[width=8cm]{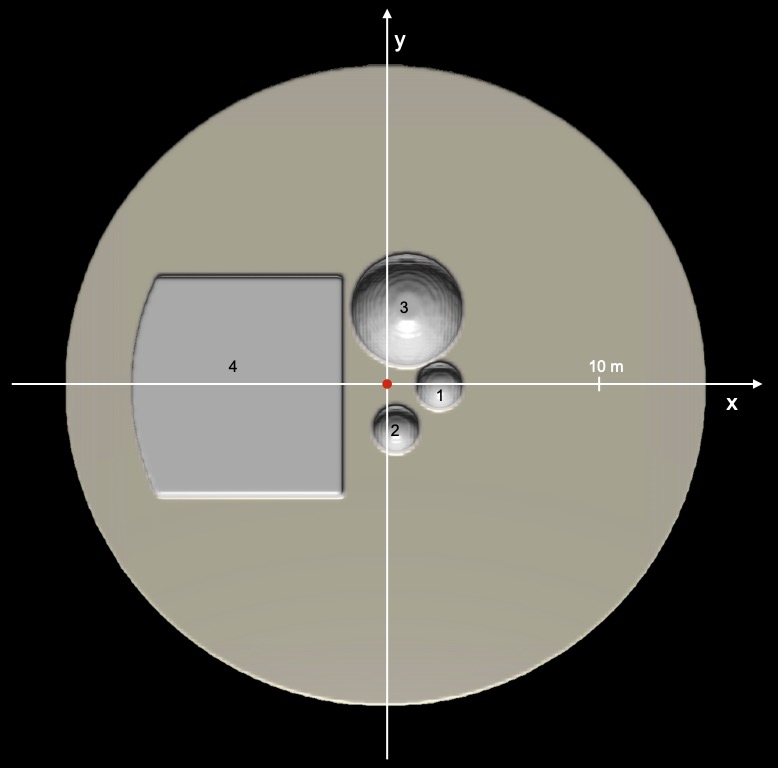}
\caption{Boulder positions and impact geometry. The impact point is in the centre, marked by the red dot. The location and dimension of the boulders (1-4) are given in Table 2.}
\label{fig:boulder_geometries}
\end{center}
\end{figure}

The positions of the boulders with respect to the impact point are shown in Fig.~\ref{fig:boulder_geometries}. The boulders have the same density as the matrix material; their dimensions are given in Table \ref{table:boulder_parameters}. 
In order to model the large subsurface Okamoto boulder (Fig.~\ref{fig:surface_snapshots}) \cite{Honda:2021}, we use a  rectangular block with the dimensions and position indicated in Table \ref{table:boulder_parameters}. The left side of this block (Fig.~\ref{fig:boulder_geometries}) has a rounded shape to avoid interference with the boundary of the overall (hemispherical) computational domain.

\FloatBarrier

\begin{table}[h]
 \centering 
    \footnotesize
	\caption{Position and properties of boulders in the simulations with heterogeneous targets.}
	\vspace{0.3cm}
	\begin{tabular}{l@{\hskip 0.1in}l@{\hskip 0.1in}l@{\hskip 0.1in}l}
    Boulder number        & Position (x, y, z) (cm) & Radius (cm)   \\
    \hline
    1             & 250, 0, 0 &   100  \\
    \hline
    2             & 50, -200, 0 &   100  \\
    \hline
    3             & 100, 350, 0 &   250  \\
    \hline
   \multirow{4}{*}{4} & $\sqrt{x^2+y^2}<900$& \multirow{4}{*}{--} \\
   
   & $x<-200$ & \\
   & $|y|<500$ & \\ 
   & $-500<z<50$ &   \\
    \hline
	\end{tabular}
	\label{table:boulder_parameters}
\end{table}
\FloatBarrier

\subsection*{Crater size computation}

In order to determine the size of the final craters, we first compute the continuum density distribution using the SPH kernel interpolation. We then define the crater radius as the point where the density at the level of the pre-impact surface falls below a critical value $\rho_{crit} = \rho_{initial}/2$, where  $\rho_{initial}$ is the initial bulk density of the homogeneous material. The final crater radius is determined as an average of the radii determined in this way along the $x$-axis and the $y$-axis. 

\section*{Crater scaling relationships}

The crater scaling relationships \citep{Holsapple1987} relate the outcome of an impact to the impact properties (e.g., impactor radius $a$, impactor density $\delta$ and impact speed $U$) by a so-called coupling parameter, which is given by $C \approx a \delta^\nu U^\mu$. Here, $\nu$ and $\mu$ are target material specific exponents. The density scaling exponent is often assumed to be $\nu \approx 0.4$ \citep{Schmidt1980}. The velocity scaling exponent, $\mu$, is one of the main scaling constants needed to extrapolate lab-scale and numerical results to other regimes of applicability \citep[e.g.,][]{Housen2011, Prieur2017,Raducan2019}. $\mu$ depends on the target material properties and its value lies between two theoretical limits: $\mu = 1/3$ if the crater formation is influenced by the impactor momentum alone and $\mu$ = 2/3 if it is influenced by the impactor energy alone. 

The most general form of the scaling relationship for crater radius, $R$, as a function of impactor properties and assuming vertical impact of a spherical impactor, is given by Eq.~\eqref{eq:pi_R}. 
%%%%%%%%%%%%%%%%
According to this equation, the scaled crater radius is proportional to
\begin{equation}
R\left(\frac{\rho}{m}\right)^{1/3} = \frac{R}{a}\left(\frac{3 \rho}{4 \pi \delta}\right)^{1/3} = D_{crat}/D_{proj}\left(\frac{3 \rho}{4 \pi \delta}\right)^{1/3}
\end{equation}

From Eq. (\ref{eq:pi_R}) it can be seen that $D_{crat}/D_{proj}$ is constant (for varying $a$) only if $g a << Y/\rho$, i.e., in the strength regime. This means that using a constant $D_{crat}/D_{proj}=10$ \citep{Bottke2020} implicitly assumes that the cratering outcome is given by purely strength regime scaling, independently of the scale of the event and the actual material strength. 

\subsection*{Calculations of the impact frequencies}
The relative impact frequencies are computed by assuming an impactor population described by a differential size distribution with single slope $q$ \citep{Jutzi2019}. We do not compute absolute numbers but rather the ratios resulting from different scaling-law parameters. Based on Eq. (\ref{eq:pi_R}), the impactor size for a given cater size is obtained which is then used as input to compute the number of expected impacts, following \cite{Jutzi2019}.

%%%%%%%%%%%%%%%%%%%%%%%%%%%%%%%%%%%%%%%%%%%
\newpage

\newpage
\section*{Data availability}
The SPH and pkdgrav simulation data generated in this study are available from the corresponding author upon reasonable request. Source data are provided with this paper.

\section*{Code availability}
A compiled version of the Bern SPH code and the pkdgrav code, as well as the necessary input files are available from the corresponding author upon reasonable request.
SPH data visualisation was produced using Vapor (www.vapor.ucar.edu).

\newpage
\bibliographystyle{naturemag}
\bibliography{bibdata}

\newpage
\section*{Acknowledgements}
The authors acknowledge funding support from the European Union's Horizon 2020 research and innovation programme under grant agreement No. 870377 (project NEO-MAPP). This work has been carried out within the framework of the NCCR PlanetS supported by the Swiss National Science Foundation under grants 51NF40$\_$182901 and 51NF40$\_$205606. P.M. Acknowledges funding support from the French space agency CNES and from the CNRS through the MITI interdisciplinary programs.

\section*{Author contribution}
M.J. performed and analysed the numerical simulations and led the research. S.D.R. helped to design the numerical study and performed the scaling-law analysis. Y.Z. and P.M. performed and analysed the SSDEM simulations. M.A. provided the SCI impact data. All authors contributed to interpretation of the results and preparation of the manuscript.

\section*{Competing interest}
The authors declare no competing interests.

\end{document}